\documentstyle[12pt]{article}
\def\be{\begin{equation}}
\def\ee{\end{equation}}
\def\bea{\begin{eqnarray}}
\def\eea{\end{eqnarray}}

\def\ba{\begin{array}}
\def\ea{\end{array}}
\def\bce{\begin{center}}
\def\ece{\end{center}}
\def\bfi{\begin{figure}}
\def\efi{\end{figure}}

\renewcommand{\thefootnote}{\fnsymbol{footnote}}

\def\ggtmn{g^{\mu\nu}}
\def\ggbmn{g_{\mu\nu}}

\def\ggtmr{g^{\mu\rho}}
\def\ggtns{g^{\nu\sigma}}
\def\ggtrs{g^{\rho\sigma}}

\def\ggttk{g^{\tau\kappa}}
\def\ggbmtt{g_{\mu}^{\;\tau}}
\def\ggbntk{g_{\nu}^{\;\kappa}}
\def\ggbmtr{g_{\mu}^{\;\rho}}
\def\ggbnts{g_{\nu}^{\;\sigma}}

\def\gggg{\sqrt g }

\def\abm{A_\mu}

\def\abn{A_\nu}

\def\abr{A_\rho}

\def\abz{A_0}

\def\abi{A_i}

\def\ftmn{F^{\mu\nu}}
\def\fbmn{F_{\mu\nu}}

\def\fbmr{F_{\mu\rho}}

\def\fbrs{F_{\rho\sigma}}
\def\fbns{F_{\nu\sigma}}

\def\fbij{F_{ij}}
\def\fbkl{F_{kl}}

\def\rbmn{R_{\mu \nu }}

\def\tbmn{T_{\mu\nu}}

\def\ttibz{T^{i}_{\; 0}}

\def\pbm{\partial_\mu}
\def\ptn{\partial^\nu}
\def\pbn{\partial_\nu}

\def\pbr{\partial_\rho}

\def\pbs{\partial_\sigma}

\def\pbi{\partial_i}

\def\pbj{\partial_j}

\def\cdbm{D_\mu}
\def\cdbk{D_k}

\def\cdbn{D_\nu}

\def\cdbi{D_i}
\def\cdbj{D_j}
\def\cdbl{D_l}

\def\tint{\int d^2\! x\,\,}
\def\thint{\int d^3\! x\,\,}

\def\etmnr{\epsilon^{\mu\nu\rho}}
\def\ebmnr{\epsilon_{\mu\nu\rho}}

\def\etij{\epsilon^{ij}}
\def\ebij{\epsilon_{ij}}

\def\ebkl{\epsilon_{kl}}
\def\ebik{\epsilon_{ik}}
\def\ebjm{\epsilon_{jm}}

\def\samp{|\phi|}

\def\half{\frac{1}{2}}

\newcommand{\np}[1]{Nucl.\ Phys.\ {\bf {#1}}}
\newcommand{\pl}[1]{Phys.\ Lett.\ {\bf {#1}}}

\newcommand{\pr}[1]{Phys.\ Rev.\ {\bf {#1}}}
\newcommand{\prl}[1]{Phys.\ Rev.\ Lett.\ {\bf {#1}}}
\newcommand{\ijmp}[1]{Int.\ J.\ Mod.\ Phys.\ {\bf {#1}}}

\newcommand{\prp}[1]{Phys.\ Rep.\ {\bf {#1}}}
\newcommand{\ap}[1]{Ann.\ Phys.\ {\bf {#1}}}

\newcommand{\spj}[1]{Sov.\ Phys.\ JETP {\bf {#1}}}
\newcommand{\sjnp}[1]{Sov.\ J.\ Nucl.\ Phys.\ {\bf {#1}}}

\catcode`\@=11
\def\maketitle{\par
 \begingroup
 \def\thefootnote{\fnsymbol{footnote}}
 \def\@makefnmark{\hbox
 to 0pt{$^{\@thefnmark}$\hss}}
 \if@twocolumn
 \twocolumn[\@maketitle]
 \else \newpage
 \global\@topnum\z@ \@maketitle \fi\thispagestyle{empty}\@thanks
 \endgroup
 \setcounter{footnote}{0}
 \let\maketitle\relax
 \let\@maketitle\relax
 \gdef\@thanks{}\gdef\@author{}\gdef\@title{}\let\thanks\relax}
\def\@maketitle{\newpage
 \null
 \hbox to\textwidth{\hfil\hbox{\begin{tabular}{r}\@preprint\end{tabular}}}
 \vskip 2em \begin{center}
 {\Large\bf \@title \par} \vskip 1.5em {\normalsize \lineskip .5em
\begin{tabular}[t]{c}\@author
 \end{tabular}\par}
 \end{center}
 \par
 \vskip 1.5em}
\def\preprint#1{\gdef\@preprint{#1}}
\def\abstract{\if@twocolumn
\section*{Abstract}
\else \normalsize
\begin{center}
{\large\bf Abstract\vspace{-.5em}\vspace{0pt}}
\end{center}
\quotation
\fi}
\def\endabstract{\if@twocolumn\else\endquotation\fi}
\catcode`\@=12

\begin{document}
\baselineskip=.285in

\preprint{KHTP-93-05\\[-2mm] SNUTP-93-100\\[-2mm] DPNU-93-46}

\title{\Large\bf Vortices in Bogomol'nyi Limit of
Einstein Maxwell\\Higgs Theory with or without External 
Sources\protect\\[1mm]\  }
\author{}
\author{\normalsize Chanju Kim\\
{\normalsize\it Center for Theoretical Physics and
Department of Physics}\\
{\normalsize\it Seoul National University,
Seoul 151-742, Korea}\\[4mm]
\normalsize Yoonbai Kim\thanks{Present address: Department of Physics, Nagoya 
University, Nagoya 461-01, Japan,
e-mail: yoonbai$@$eken.phys.nagoya-u.ac.jp}\\
{\normalsize\it Research Institute for Basic Sciences and Department of Physics
}\\
{\normalsize\it Kyung Hee University,
\normalsize\it Seoul 130-701, Korea}}
\date{}
\maketitle

\renewcommand{\theequation}{\thesection.\arabic{equation}}

\def\gatij{\gamma^{ij}}
\def\gabij{\gamma_{ij}}

\begin{center}
{\large\bf Abstract}\\[3mm]
\end{center}
\indent\indent
The Abelian Higgs model with or without external particles is considered 
in curved space.
Using the dual transformation, we rewrite the model in terms of dual gauge
fields and derive the Bogomol'nyi-type bound. 
We find all possible cylindrically symmetric vortex solutions and 
vortex-particle composites by examining the Einstein equations and the 
first-order Bogomol'nyi equations. The underlying spatial manifold of these
objects comprises a cylinder asymptotically and a two sphere in addition to
the well-known cone.

\newpage

\pagenumbering{arabic}
\thispagestyle{plain}
\setcounter{section}{1}
\begin{center}\section*{\large\bf I. Introduction}\end{center}
\indent\indent A classic example of solitons in field theories is the
topologically stable vortex solution in Abelian Higgs model
\cite{Abr,NO}. Such vortex-like excitations have been used to describe 
the system of flux-tubes in type-II superconductors \cite{Abr}.
These objects have played considerable roles also as cosmic strings.
For example, phase transitions in early universe give rise to string-like
defects and they generate the density fluctuation which leads to
a possible explanation of galaxy formation \cite{Vil,GHV}. 

In this paper, we  consider $U(1)$ local strings of Einstein Maxwell Higgs 
theory. However, the model may be too complex to handle analytically if one 
considers 
arbitrary shaped cosmic strings coupled to gravity. Thus one may adopt local
cosmic strings which are infinitely straight in one direction and remain
in equilibrium, which describes
an idealized but a physically relevant situation.
This assumption simplifies the system to that of Nielsen-Olesen vortices 
coupled to Einstein gravity in (2+1) spacetime dimensions, and the zero size 
limit
of such objects can be identified with point particles on a plane \cite{DJH}.
Then (2+1) dimensional Einstein Maxwell Higgs theory with or without the 
coupling of external point particles is of our interest, and we examine possible
vortex configurations whose centers point particles may stick to.
Furthermore, if we choose a specific form of scalar potential, static 
local vortex configurations can be solutions of a first-order differential
equations, which satisfies second-order Euler-Lagrange equation 
automatically \cite{Bog,HKP}. 
These Bogomol'nyi equations have been established in the theory coupled to 
gravity despite the difficulty
of constructing the gravitational energy \cite{LCG,Val,GORS}.
For such vortex solutions to this Bogomol'nyi equation, we show
that they are exact solutions of Einstein equation and that the cosmological
constant is zero for a general stationary metric and arbitrary distribution
of point particles. Specifically, the Bogomol'nyi-type bound is saturated when
the metric is static \cite{LCG}.

Another way of understanding the role of vortices in $U(1)$ gauge theory is to
recapitulate it in dual formulation and elicit the physically relevant aspects
such as phase transitions induced by vortices and classical dynamics
of vortices \cite{Sav,KL}. Here we derive the dual-transformed version of
Einstein Maxwell Higgs theory in (2+1) dimensions, and show that the
Bogomol'nyi-type bound can be attained within this formulation. It is found
that the extended objects corresponding to the vortices in the original
theory do not carry dual magnetic flux but dual electric field, and
the energy which is proportional to Euler invariant is expressed by the sum of 
the spatial integral of electrostatic potential
and the total mass of point particles in the Bogomol'nyi limit.

Bogomol'nyi equations being set up,
we investigate cylindrically symmetric configurations and find all possible
vortex-particle solutions by explicitly proving their existence.
Interestingly, there are solutions whose two manifold constitutes
a cylinder in its asymptotic form or a two sphere, in addition 
to those solutions whose two manifold constitutes a cone.

The rest of this paper is organized as following. In section 2 we introduce the
model and derive Bogomol'nyi-type bound
in a more detailed way by solving Einstein equations under general stationary
metric. Section 3 is devoted to the dual transformation of
Einstein Maxwell Higgs theory within path-integral formalism and to the study
of various aspects of the dual formulation, including the derivation of 
Bogomol'nyi-type bound. In section 4 we analyze cylindrically symmetric 
solutions and their global geometrical structures.
We conclude in section 5 with some remarks.

\setcounter{section}{2}
\setcounter{equation}{0}
\bce\section*{\large\bf II. Einstein Maxwell Higgs Theory and Bogomol'nyi Bound }
\ece
\indent\indent We consider the system composed of vortices in Abelian Higgs
model and the massive point particles in the presence of gravity. 
Einstein Maxwell Higgs theory
coupled to a set of $N$ external particles
in (2+1) spacetime dimensions is described by the action
\bea \label{action}
S&=&S_{gravity}+S_{matter}+S_{point\; particle}\nonumber\\
&=&\thint\!\!\gggg \left\{\, -\,\frac{1}{16\pi G} (R+2\Lambda)
-\frac{1}{4} \ggtmr \ggtns \fbmn \fbrs 
+ \,\half\ggtmn \overline{\cdbm \phi} \cdbn \phi -
V(\samp)\,\right\}\nonumber\\
&&+\sum_{a=1}^{N}m_{a}\int_{-\infty}^{\infty}\! ds\sqrt{\ggtmn
\frac{dx^{a}_{\mu}}{ds}\frac{dx^{a}_{\nu}}{ds}}\; ,
\eea
where $\phi=e^{i\Omega}\samp$, $\cdbm\phi=(\pbm-ie\abm)\phi$, 
and $\Lambda$ the cosmological constant.

Equations of motion read
\bea
\frac{1}{\gggg }\pbn \left(\gggg \ftmn\right) & = &e j^{\mu }
\label{Maxeq}\\
\frac{1}{2\gggg} \cdbm \left(\gggg \ggtmn \cdbn \phi \right)
&=&-\frac{\partial V}{\partial\bar{\phi}} 
\label{scaEQ} \\
\rbmn - \half \ggbmn (R+2\Lambda )
& = & 8\pi G \tbmn , \label{EinEq}
\eea
where $j^{\mu}$ is the conserved U(1) current,
\be \label{current}
j^\mu = -\frac{i}{2}\ggtmn \left(\, \bar{\phi} \cdbn \phi
-\overline{\cdbn \phi} \phi \,\right),
\ee
and $\tbmn$ is the energy-momentum tensor,
\bea \label{enemom}
\tbmn&=&- \ggtrs \fbmr \fbns + \half \left(\; \overline{\cdbm \phi}
\cdbn \phi + \overline{\cdbn \phi} \cdbm \phi \; \right)
-\ggbmn{\cal L}_{matter}\nonumber\\
&&+\frac{1}{\gggg}\sum_{a=1}^{N}m_{a}\int_{-\infty}^{\infty}ds_{a}
\delta(x_{a}(s_{a})-x)\frac{dx^{a}_{\mu}}{ds_{a}}\frac{dx^{a}_{\nu}}{ds_{a}}.
\eea

We are interested in static soliton solutions of the equations of motion, 
specifically the Nielsen-Olesen vortices which are electrically-neutral
and are characterized by the magnetic flux
\be
\Phi=-\tint \frac{1}{2}\etij \fbij, 
\ee
where $\etij$ two dimensional Levi-Civita tensor density of 
$\epsilon^{12}=\epsilon_{12}=1$.
With an appropriate choice of the potential $V(\psi)$, 
let us look for multivortex-particle configurations in curved spacetime 
which satisfy Bogomol'nyi-type equations. If we assume that the vortices satisfy
such equations, Einstein equations will turn out to be 
solved under the general stationary metric of the form
\be
ds^2 = N^2 (dt + K_i dx^i )^2 - \gamma_{ij} dx^i dx^j ,
\ee
where the functions $N(x), K_i (x), \;{\rm and}\;
\gamma_{ij}(x),\;  (i,j=1,2)$, are independent of time.

Under $\abz=0$ gauge condition, $0i$-components of energy-momentum tensor 
$\ttibz$ 
vanish, so the corresponding solution of Einstein equations is
\be \label{kappa}
K_{ij}=\kappa\frac{\ebij}{N^{3}\!\!\sqrt{\gamma}},
\ee
where 
$K_{ij}
=\pbi K_{j}-\pbj K_{i}$, 
and $\kappa$ is a undetermined constant.
Taking $N=1$ and rearranging the terms in the spatial integration of 
the $00$-component of energy-momentum tensor, we have 
\bea\label{tzz}
\lefteqn{\tint\!\!\gggg \; T_{00}}\nonumber\\
&=&\tint\!\!\sqrt{\gamma}\,\Biggl\{\frac{1}{4}\gamma^{ik}\gamma^{jl}(\fbij\mp
\sqrt{\gamma}\ebij\sqrt{2V})(\fbkl\mp\sqrt{\gamma}\ebkl\sqrt{2V})\nonumber\\
&&\hspace{19mm}+\frac{1}{4}\gamma^{ij}(\overline{\cdbi\phi\mp i\sqrt{\gamma}\ebik
\gamma^{kl}\cdbl\phi})(\cdbj\phi\mp i\sqrt{\gamma}\ebjm\gamma^{mn}D_{n}\phi)
\nonumber\\
&&\hspace{19mm}\pm\frac{\etij}{\sqrt{\gamma}}\fbij\Bigl(\sqrt{2V}-\frac{e}{2}(\samp^{2}
-v^{2})\Bigr)\Biggr\}\\
&&\pm\frac{ev^{2}}{2}\Phi+\sum^{N}_{a=1}m_{a}\nonumber\\
&&\pm\frac{1}{2e}\tint\pbi(\sqrt{\gamma}\etij j_{j}).\nonumber
\eea
Now we choose the scalar potential as
\be\label{sp}
V=\frac{e^{2}}{8}(\samp^{2}-v^{2})^{2},
\ee
then the first two terms in the braces of Eq. (\ref{tzz}) are nonnegative 
definite and we obtain
the so called Bogomol'nyi bound. The bound is attained for configurations
satisfying the Bogomol'nyi
equations
\be\label{bog1}
\cdbi\phi\mp i \sqrt{\gamma}\ebij\gamma^{jk}\cdbk\phi=0
\ee
\be\label{bog2}
\frac{1}{2}\frac{\etij}{\sqrt{\gamma}}\fbij=\pm\sqrt{2V}.
\ee
In this Bogomol'nyi limit, the $ij$-components of 
energy-momentum tensor vanish and then
Einstein equations force both an integration constant $\kappa$ and the cosmological constant
$\Lambda$ to be zero. Thus we have no off-diagonal components of metric $K_{i}$ 
up to a gauge. 
The 00-component of Einstein equations 
relates the topological quantities, {\it i.e.} the
geometric part gives the Euler number and the matter part, 
the sum of magnetic flux from vortices
and total mass from point particles 
\be \label{Eul}
-\frac{1}{16\pi G} \tint\!\!\sqrt{\gamma}\;\, { }^{2}\! R=\frac{v^{2}}{2}|e\Phi|
+\sum_{a=1}^{N}m_{a}.
\ee

After eliminating the gauge field $\abi$ by use of a Bogomol'nyi equation (\ref{bog1})
\be \label{abi}
\abi=\frac{1}{e}(\pbi\Omega\mp\sqrt{\gamma}\,\ebij\gamma^{jk}\partial_{k} \ln\samp),
\ee
and fixing the gauge for the spatial components of metric
\be
\gamma_{ij}=-\delta_{ij} b(x^{i}),
\ee
we solve 00-component of Einstein equations
\be\label{metr}
b(x^{i})=e^{h(\tilde{z})+\bar{h}(\bar{\tilde{z}})}
\left(\frac{f^2e^{-(f^2-1)}}{\prod^{n}_{p=1}|\tilde{z}-\tilde{z}_{p}|^{2}
\prod_{a=1}^{N}|\tilde{z}-\tilde{z}_{a}|^{2\tilde{m}_{a}}}\right)^{\tilde{G}}, 
\ee
where $h(\tilde{z})\; (\bar{h}(\bar{\tilde{z}}_{p}))$ is a 
holomorphic (antiholomorphic) function and 
the tilded variables are dimensionless quantities
\be
\tilde{z}=\tilde{x}^{1}+i\tilde{x}^{2}=ev(x^{1}+i x^{2}),\; f=\frac{\samp}{v},\;
\tilde{G}=4\pi Gv^{2},\; \tilde{m}_{a}=\frac{m_{a}}{\pi v^{2}}.
\ee
If the spacetime of the vacuum (no particle $(N=0)$ and no vortex $(n=0)$)
is to be Minkowski, the harmonic function should be chosen to be zero, 
$h(\tilde{z})+\bar{h}(\bar{\tilde{z}})=0$. 
Substituting Eq.(\ref{metr}) and Eq.(\ref{abi}) into Eq.(\ref{bog2}), we obtain
a single equation for the amplitude of Higgs field
\be\label{beq}
\tilde{\partial}^{2}\ln f^2=e^{h+\bar h}
\left(\frac{f^2e^{-(f^2-1)}}{\prod^{n}_{p=1}|\tilde{z}-\tilde{z}_{p}|^{2}
\prod_{a=1}^{N}|\tilde{z}-\tilde{z}_{a}|^{2\tilde{m}_{a}}}\right)^{\tilde{G}} 
(f^2-1)+4\pi\sum_{p=1}^{n}\delta^{(2)}(\tilde{z}-\tilde{z}_{p}),
\ee
where $\tilde{\partial}^{2}$ is a Laplacian in flat two-dimensional space.
Returning to the expression of Euler number in Eq.(\ref{Eul}) and inserting
Eq.(\ref{metr}) into it, we have
\bea\label{eule}
\lefteqn{\frac{1}{16\pi G}\tint\!\!\sqrt{\gamma}\;{}^{2}\! R,} \nonumber\\
&=&\frac{v^{2}}{4}\left\{\tint \partial^2 \ln\Bigl(\prod_{p=1}^{n}|z-z_{p}|^{2}
\prod_{a=1}^{N}|z-z_{a}|^{\frac{2m_{a}}{\pi v^{2}}}\Bigr) 
-\tint \partial^2 \ln \samp^2 + \tint \partial^2\frac{\samp^2}{v^{2}}\right\} .\nonumber\\
&&
\eea
{}From the above expression one may notice that, in addition to
the contribution from the first two terms for $n\neq0$ or $m_{a}\neq0$, there can exist 
contribution from the second term if the Bogomol'nyi
equation (\ref{beq}) contains the finite energy solution which behaves as
$\samp \sim |\vec{x}|^{-\varepsilon} \; (\varepsilon > 0) $
for large $|\vec{x}|$. We shall show that it is indeed the case and present 
the detailed analysis for the existence of
such solutions in section 4.

\setcounter{section}{3}
\setcounter{equation}{0}
\begin{center}\section*{\large\bf III. Dual Formulation}
\end{center}
\indent\indent
In this section we reformulate the 
theory by use of the dual transformation and
re-derive the Bogomol'nyi limit in this formulation.
The path integral for the system we are considering is given by
\bea \label{Dual}
Z&=&\left< F|e^{iHT}|I\right>\nonumber\\
&=&\int [d\ggbmn ][dA_{\mu}][\samp d\samp ][d\Omega ]\nonumber\\
&&\times\exp i\Biggl\{\thint\!\gggg\left[ -\frac{1}{16\pi G}(R+2\Lambda )
-\frac{1}{4}\ggtmr\ggtns\fbmn\fbrs+\frac{1}{2}\ggtmn\pbm\samp
\pbn\samp\right.\nonumber\\
&&\hspace{3.45cm}\left. +\frac{1}{2}\ggtmn\samp^{2}(\pbm\Omega-eA_{\mu})
(\pbn\Omega-e\abn)-V(\samp )\right]\nonumber\\
&&\hspace{1.65cm} +\sum_{a=1}^{N}m_{a}\int_{-\infty}^{\infty}ds\sqrt{\ggtmn
\frac{dx^{a}_{\mu}}{ds}\frac{dx^{a}_{\nu}}{ds}}\;\Biggr\} .
\eea

Introducing an auxiliary vector field $C_{\mu}$, we rewrite the
interaction term between scalar field and gauge field as the following
\bea \label{Lin}
\lefteqn{\exp \left\{ i\thint\!\!\gggg\;
\frac{1}{2}\ggtmn\samp^{2}(\pbm\Omega-eA_{\mu})
   (\pbn\Omega-e\abn)\right\} }\nonumber\\
&=&\int [dC_{\mu}]\prod_{x}\frac{g^{\frac{1}{4}}}{\samp^{3}}\exp i\thint
\!\!\gggg\left\{ -\frac{\ggtmn}{2\samp^{2}}
C_{\mu}C_{\nu}+\ggtmn C_{\mu}(\pbn\Omega -e\abn )\right\}.
\eea
Classically the auxiliary field $C_{\mu}$ is nothing but the conserved $U(1)$
current.

Since the theory contains vortex and antivortex configurations,
the phase of the scalar field need not be single-valued and then can be
split into two parts
\be
\Omega(t,\vec{x})=\Theta(t,\vec{x})+\eta(t,\vec{x}).
\ee
The first term $\Theta$ which describes a configuration of
vortices and antivortices is defined by a multi-valued function
\be \label{Theta}
\Theta(t,\vec{x})
  =\sum_{p}(\mp)\tan^{-1}\frac{x^{2}-x^{2}_{p}(t)}{x^{1}-x^{1}_{p}(t)},
\ee
where $(x^{1}_{p}(t), x^{2}_{p}(t))$ is the position of a (anti-)vortex and 
$(\mp)$ denotes $-1$ for vortex and 1 for anti-vortex.
The single-valued function $\eta$ represents the
fluctuation around a given vortex sector. 
Hence the path integral measure is divided into two contributions
\be
[d\Omega]=[d\Theta][d\eta],
\ee
{\it i.e.} the one for the sum over single-valued fluctuation around a
given configuration of vortices and the other
for that over all possible configurations of vortices, including
the annihilation and creation of vortex-antivortex pairs.

After $\eta$-integration,
\be \label{CuCon}
\int [d\eta] \exp \left\{ i\thint\!\!\gggg\;\ggtmn C_{\mu}\pbn\eta
\right\}\approx \frac{1}{\gggg}\delta(\nabla_{\mu} C^{\mu}),
\ee
we can rewrite a part of the path integral by introducing dual gauge
field $H_{\mu}$,
\be \label{Meas}
\int [dC_{\mu}]\frac{1}{\gggg}\delta (\nabla_{\mu}C^{\mu})\cdots
=\int [dH_{\mu}][dC_{\mu}]\delta(\gggg\, C^{\mu}-\frac{1}{e}\etmnr\pbn H_{\rho})\cdots\, .
\ee
Substituting Eq.(\ref{Meas}) into the path integral in Eq.(\ref{Lin}) and integrating out the
auxiliary field $C_{\mu}$, we obtain an effective theory described by the action
\bea
S^{'}&=&\thint\!\!\gggg\Biggl\{-\frac{1}{16\pi G}(R+2\Lambda )
-\frac{1}{2}\ggtmn\tilde{F}_{\mu}\tilde{F}_{\nu}+\frac{1}{2}\ggtmn
\pbm\samp\pbn\samp -V(\samp)\nonumber\\
&&\hspace{1.9cm}-\frac{1}{4e^{2}\samp^{2}}\ggtmr\ggtns H_{\mu\nu}H_{\rho\sigma}
+\frac{1}{2}\frac{\etmnr}{\gggg}H_{\mu\nu}(\frac{1}{e}\pbr\Theta -\abr)
\Biggr\}\\
&& +\sum_{a=1}^{N}m_{a}\int_{-\infty}^{\infty}ds\sqrt{\ggtmn
\frac{dx^{a}_{\mu}}{ds}\frac{dx^{a}_{\nu}}{ds}} ,\nonumber
\eea
where $H_{\mu\nu}=\pbm H_{\nu}-\pbn H_{\mu}$ and 
$\displaystyle{\tilde{F}^{\mu}=\frac{\etmnr}{2\!\gggg}F_{\nu\rho}}$,
the dual of field strength tensor. For performing $\abm$
integration, let us regard $\abm$ and $\tilde{F}^{\mu}$ as independent variables and rewrite the
path integral measure by introducing another auxiliary field $N_{\mu}$
\bea \label{condi}
\lefteqn{ \int [d\abm ][d\tilde{F}^{\mu}]\,\delta (\tilde{F}^{\mu}-
\frac{\etmnr}{\gggg}\pbn\abr) \cdots }
\nonumber\\
&=&\int [d\abm ][d\tilde{F}^{\mu}][d N_{\mu}]\prod_{x}\!\gggg\,
\exp \left\{ i\thint\!\!\gggg\, N_{\mu}(\tilde{F}^{\mu}-\frac{\etmnr}{\gggg}
\pbn\abr)\right\}\cdots \, .
\eea
Putting the above equation (\ref{condi}) into the path integral and then integrating
$\tilde{F}^{\mu}$ and $\abm$ fields out, we have
\bea\label{equal}
\lefteqn{\int [d\abm ]\exp i\thint\!\gggg\left\{
-\frac{1}{2}\ggbmn\tilde{F}^{\mu}\tilde{F}^{\nu} +H_{\mu}\tilde{F}^{\mu}\right\}}
\nonumber\\
&=&\!\int [dN_{\mu}]\prod_{x}\! g^{\frac{1}{4}}\,\delta(\etmnr\pbn N_{\rho})
\exp i\!\thint\!\!\gggg\;\frac{\ggtmn}{2}(H_{\mu}+N_{\mu})(H_{\nu}+N_{\nu})\, .
\eea
The delta functional in Eq.(\ref{equal}) implies
$N_{\mu}=-\pbm \chi$ for a single-valued scalar field $\chi$.
Finally the path integral of dual transformed theory becomes
\be
Z=\int [gd\ggbmn][d H_{\mu}][\samp^{-2}d\samp][d\Theta ][d\chi] \,\exp\left\{
iS_{D}\right\},
\ee
where the action of dual transformed theory is
\bea
S_{D}&=&\thint\!\!\gggg\left\{-\frac{1}{16\pi G}(R+2\Lambda )+\frac{1}{2}\ggtmn
\pbm\samp\pbn\samp -V(\samp)
\right.\nonumber\\
&&\left.\hspace{18mm}
-\frac{1}{4e^{2}\samp^{2}}\ggtmr\ggtns H_{\mu\nu}H_{\rho\sigma}
+\frac{1}{2e}\frac{\etmnr}{\gggg}H_{\mu\nu}
\pbr\Theta +\frac{1}{2}\ggtmn(H_{\mu}-\pbm\chi)
(H_{\nu}-\pbn\chi )\right\}\nonumber\\
&&\left. +\sum_{a=1}^{N}m_{a}\int_{-\infty}^{\infty}ds\sqrt{\ggtmn
\frac{dx^{a}_{\mu}}{ds}\frac{dx^{a}_{\nu}}{ds}}\right. .
\eea
In Higgs phase this dual-transformed theory describes a gauge boson of mass $ev$
and a neutral Higgs. Gauge coupling $e$ is inversely multiplied to the interaction
term between the gauge field and the Higgs field, which looks like the strong 
coupling expansion being done. However, when Higgs effects are important, one must
take into account the nonpolynomial interaction in the Maxwell-like term and the 
Jacobian in the measure of the Higgs field. Though the classical gravity is not affected by the 
dual transformation, the Jacobian in the measure of gravitational field is introduced
and this induced Jacobian factor depends on both the gauge dynamics and the 
dimension of spacetime \cite{CCK}.

The Euler-Lagrange equations are
\be \label{DEq1}
\frac{1}{\gggg}\pbm (\gggg\ggtmn\pbn\samp)=\frac{1}{2e^{2}\samp^{3}}
\ggtmr\ggtns H_{\mu\nu}H_{\rho\sigma}-\frac{dV}{d\samp}
\ee
\be \label{DEq2}
\frac{1}{\gggg}\pbn\left(\gggg\frac{H^{\mu\nu}}{e^{2}\samp^{2}}\right)
-\ggtmn(H_{\nu}-\pbn\chi)=\frac{1}{e}\frac{\etmnr}{\gggg}
\pbn\pbr\Theta
\ee
\be \label{DEq3}
R_{\mu\nu}-\frac{1}{2}\ggbmn(R+2\Lambda)=T^{D}_{\mu\nu},
\ee
where the energy-momentum tensor of dual-transformed theory is
\bea \label{DEM}
T^{D}_{\mu\nu}&=&\frac{1}{4e^{2}\samp^{2}}\ggtrs(\ggbmn\ggttk-4\ggbmtt\ggbntk)
H_{\rho\tau}H_{\sigma\kappa}
+(\ggbmtr\ggbnts-\frac{1}{2}\ggbmn\ggtrs)(H_{\rho}-\pbr\chi)(H_{\sigma}-\pbs\chi)
\nonumber\\
&&+(\ggbmtr\ggbnts-\frac{1}{2}\ggbmn\ggtrs)\pbr\samp\pbs\samp+\ggbmn V\\
&&+\frac{1}{\gggg}\sum_{a=1}^{N}m_{a}
\int_{-\infty}^{\infty}\! ds_{a}\, \delta(x_{a}(s_{a})-x)
\frac{dx^{a}_{\mu}}{ds_{a}}\frac{dx^{a}_{\nu}}{ds_{a}}.\nonumber
\eea

{}From now on let us look for the counterpart of the vortices in the
original theory through the derivation of Bogomol'nyi-type bound
in the dual-transformed theory.
{}From Eq.(\ref{Lin}) and Eq.(\ref{Meas}), we notice that the classical
configurations of both formulations are related by
\be
\samp^{2}(\pbm\Omega-e\abm)=\frac{1}{e}\gggg\ebmnr\ptn H^{\rho}.
\ee
Together with Eq.(\ref{DEq2}), we obtain
\be \label{Drel}
\tilde{F}^{\mu}=-\ggtmn(H_{\nu}-\pbn\chi).
\ee
For the static counterpart of vortices, Eq.(\ref{Drel}) 
and the condition that static
vortices do not carry electric field
imply  
\be \label{Drei}
\tilde{F}_{i}=H_{i}-K_{i}H_{0}-\pbi\chi=0.
\ee
These solitons are characterized by dual electrostatic potential, when $N=1$
\be
\Phi_{D}=\tint\!\! \sqrt{\gamma}\,H_{0}.
\ee
The spatial components of Eq.(\ref{DEq2}) are solved with the help of Eq.(\ref{Drei}),
\be \label{kij}
K^{ij}=\kappa_{D}e^{2}\frac{\etij\samp^{2}}{N\sqrt{\gamma}H_{0}},
\ee
where $\kappa_{D}$ is a constant. Inserting this result into $0i$-components
of Einstein equations (\ref{DEq3}), we obtain
\be\label{ka}
\kappa_{D}=0 \;\;\;\mbox{or}\;\;\; \frac{N^{2}\samp^{2}}{2H_{0}}-8\pi GH_{0}=C,
\ee
where $C$ is an integration constant. Since the vortex solutions of our interest
do not satisfy the second condition in the sequel, we take the first one from now on. 
Then $T^{D\, i}_{\;\;\;\;\; 0}$ vanishes, and the vortices in dual formulation are also
spinless as expected.

The Bogomol'nyi-type bound of the dual-transformed Einstein Maxwell Higgs theory is obtained as follows.
\bea
\lefteqn{\tint\!\!\sqrt{\gamma}\, T^{D}_{00}}\nonumber\\
&=&\tint\!\!\sqrt{\gamma}\Biggl\{\frac{1}{2e^{2}\samp^{2}}\Bigl[\gamma^{ij}\pbi H_{0}\pbj H_{0}
+\frac{N^{2}}{2}\gamma^{ij}\gamma^{kl}(\tilde{H}_{ik}+K_{ik}H_{0})(\tilde{H}_{jl}
+K_{jl}H_{0})\Bigr]\nonumber\\
&&\hspace{19mm}+\frac{1}{2}(H_{0}^{2}+N^{2}\gamma^{ij}\tilde{H}_{i}\tilde{H}_{j}
)
\frac{N^{2}}{2}\gamma^{ij}\pbi\samp\pbj\samp+N^{2}V+
\frac{1}{\sqrt{\gamma}}\sum^{N}_{a=1}m_{a}\delta^{(2)}(\vec{x}-\vec{x}_{a})\Biggr\}\nonumber\\
&=&\tint\!\!\sqrt{\gamma}\Biggl\{\frac{\gamma^{ij}}{2e^{2}\samp^{2}}\pbi\left(H_{0}\pm
\frac{e}{2}(\samp^{2}-v^{2})\right)\pbj\left( H_{0}\pm\frac{e}{2}
(\samp^{2}-v^{2})\right)+\frac{1}{2}(H_{0}\mp\sqrt{2V}\;)^{2}\nonumber\\
&&\hspace{19mm}\pm H_{0}\left(\sqrt{2V}+\frac{e}{2}(\samp^{2}-v^{2})\right)\\
&&\hspace{19mm}\pm\frac{\samp^{2}}{2\sqrt{\gamma}}\biggl[\pbi\Bigl(\sqrt{\gamma}
\gamma^{ij}\frac{\pbj H_{0}}{e\samp^{2}}\Bigr)-e\sqrt{\gamma}H_{0}-\frac{\etij}{\sqrt{\gamma}}
\pbi\pbj\Theta\biggr]\Biggr\}\nonumber\\
&&\pm\frac{ev^{2}}{2}\Phi_{D}+\sum^{N}_{a=1}m_{a}\nonumber\\
&&\mp\tint\pbi\bigl(\sqrt{\gamma}\gamma^{ij}\frac{1}{2e}\pbj H_{0}\bigr).\nonumber
\eea
To get the last expression we have set $N=1$. Hence, if we
choose the scalar potential as Eq.(\ref{sp}), $H_{0}$ becomes
\be
H_{0}=\pm\frac{1}{2}(v^{2}-\samp^{2}).
\ee
Since $T^{ij}_{D}=0$ in this Bogomol'nyi limit and $K^{ij}=0$ from
Eq.(\ref{kij}) and Eq.(\ref{ka}), $ij$-components of Einstein equations are solved by
$\Lambda=0$. The $00$-component of Einstein equations is exactly solved as that of 
the original theory and 
the solution is Eq.(\ref{metr}). Substituting this result into 
Gauss' law (\ref{DEq2}), we obtain a single Bogomol'nyi equation, 
Eq.(\ref{beq}).

\setcounter{section}{4}
\setcounter{equation}{0}
\begin{center}\section*{\large\bf IV. Cylindrically Symmetric
Solution}\end{center}
\indent\indent 
At the outset of our consideration in Einstein Maxwell Higgs model, we had ten
second-order differential equations (\ref{Maxeq})$\sim$(\ref{EinEq}) (or
(\ref{DEq1})$\sim$(\ref{DEq3})) for scalar, gauge, and gravitational fields
even though the system was introduced in (2+1) dimensions after the reduction of
a dimension along $z$-axis. However, once we have limited our interest
to the static vortex solutions under a specific $\phi^{4}$ scalar potential
which saturate the Bogomol'nyi-type bound, three components of the gauge field and
six components of metric tensor have been expressed by the scalar field
and now there remains only one second-order equation (\ref{beq}) to solve. 
In this section let us look for regular and finite-energy vortex solutions of the Bogomol'nyi
equation of which the base manifold is smooth except for the positions
where the point particles lie. Let $\Sigma$ be the spatial part of the (2+1)
dimensional spacetime. It will be seen that there exist solutions
such that the space $\Sigma$ is a cone or a cylinder asymptotically, or a two 
sphere.

For the sake of tractability while keeping the main physical properties, let us
begin with examining the cylindrically symmetric solutions of Bogomol'nyi 
equation
(\ref{beq}). The metric which is compatible with cylindrically symmetric 
configurations is of the form
\be
ds^{2}=dt^{2}-\frac{1}{(ev)^{2}}b(r)(dr^{2}+r^{2}d\theta^{2}),
\ee
where $r=ev\sqrt{x^{i}x^{i}}$ and $0\leq\theta<2\pi$.
Since we consider the regular static solutions of Einstein
equations, the space $\Sigma$ described by metric $b(r)$ is smooth except 
for the points where there are massive point particles. 
The global geometry of $\Sigma$ may be 
characterized by the area $\displaystyle{{\cal A}=\frac{2\pi}{e^{2}v^{2}}
\int^{\infty}_{0}\!dr\, r\,b(r)}$, the radial distance from the origin 
$\displaystyle{\rho(r)=\frac{1}{ev}\int^{r}_{0}\!dr^{'}\sqrt{b(r^{'})}}$, and 
the circumference $\displaystyle{l(r)=\frac{2\pi}{ev}r\sqrt{b(r)}}$.

With the aid of gauge
transformation, any static cylindrically symmetric field configuration can be brought
into the following ansatz
\be \label{ans}
\phi=vf(r)e^{in\theta}\equiv v\,\exp\Bigl(\frac{u(r)}{2}+in\theta\Bigr).
\ee
Substituting the ansatz into the spatial integral of 00-component
of the Einstein equations (\ref{EinEq}) or (\ref{DEq3}), we obtain
\bea\label{Den}
\lefteqn{\frac{1}{16\pi G}\tint\!\!\sqrt{\gamma}\;{}^{2}\! R}&&\nonumber\\
&=&\frac{2\pi}{e^{2}}\int_{0}^{\infty}dr\,r\left\{
\biggl(\frac{de^{\frac{u}{2}}}{dr}\biggr)^{2}+\frac{F}{4r^{2\tilde{G}(n+\tilde{M})}}
e^{-\tilde{G}(e^{u}-u-1)}(e^{u}-1)^{2}\right\}+\tilde{M},
\eea
where $F$ is the harmonic function factor in Eq. (\ref{metr}) which is reduced
to a constant in the cylindrically symmetric case and 
$\displaystyle{\tilde{M}=\sum_{a=1}^{N}\tilde{m}_{a}}$ is the total mass
of point particles which are now superimposed at the origin.

For nonsingular solutions, $f=|\phi|/v$ should behave as
\be \label{bczero}
f\sim f_0r^n
\ee
near the origin. Further conditions are obtained by the requirement that Eq.
(\ref{Den}) is finite. Analyzing the behavior of the integrand of Eq.
(\ref{Den}) near the origin by use of Eq. (\ref{bczero}), we find a condition 
that 
\be \label{mcondition}
\tilde{G}\tilde{M}<1\,,
\ee
for finite energy solutions to exist except the trivial one $u=0$; analysis 
in the asymptotic region gives the boundary condition for $u$,
\be \label{bcinfty}
u(r=\infty)=\left\{\begin{array}{ll}
            \mbox{0 or $-\infty$}, &\mbox{if $0<\tilde{G}(n+\tilde{M})\le1$}\\
            \mbox{arbitrary number between $-\infty$ and 0,} 
             &\mbox{if $\tilde{G}(n+\tilde{M})>1$.}
            \end{array}\right.
\ee
which is enlarged in compared with that in flat space-time.
Now let us examine soliton solutions of the Bogomol'nyi equation for 
the cases of $e^{u(r=\infty)/2}\neq0$ and $e^{u(r=\infty)/2}=0$ separately. 

\vspace{5mm}
\subsection*{\normalsize\bf (a) $e^{u(r=\infty)/2}\neq0$}
\ \indent At first we consider the case that the scalar field $\displaystyle{e^{u/2}}$ does not
vanish at spatial infinity.  
As we see in Eq. (\ref{bcinfty}),
when $\tilde{G}(n+\tilde{M})$ is smaller than one
the boundary value of $u$ at $r=\infty$ has to be zero 
for finite-energy solutions. 
In this case, it is convenient to introduce a variable $R$ such that
\be\label{Var2}
R=\frac{r^{1-\tilde{G}(n+\tilde{M})}}{1-\tilde{G}(n+\tilde{M})}\,.
\ee
Then the Bogomol'nyi equation (\ref{beq}) is rewritten as
\be \label{New1}
\frac{d^{2}u}{dR^{2}}=-\frac{dV_{eff}}{du}-\frac{1}{R}\frac{du}{dR},
\ee
where $V_{eff}$ is defined by
\be \label{veff}
V_{eff}=\displaystyle{\frac{F}{\tilde{G}}\exp[-\tilde{G}(e^{u}-u-1)]}\,,
\ee
which is an increasing function of $u$ 
for $-\infty<u\leq0$ and has a maximum at $u=0$. (See Fig. 1.)
If we interpret $u$ as a particle position and $R$ as time, Eq. (\ref{New1})
is nothing but the Newton's 
equation for a particle of unit mass moving in a potential $V_{eff}$ and
subject to a friction. The particle also receives an impact at $R=0$ 
from the delta function term in Eq.(\ref{beq}). 
When $n=0$,  $u=0$ is the unique solution 
and it describes two dimensional flat space when there is no
particle and a cone when there is a massive particle at the origin. 
When $n\neq0$, we now show that there always exists a finite energy solution
whose base manifold $\Sigma$ 
is a cone asymptotically. For this we have to show that,
if we suitably choose the initial parameter $f_0$ in Eq. (\ref{bczero}), we 
can obtain the motion of the hypothetical particle such that it starts at 
negative infinity with the initial velocity given by Eq. (\ref{bczero}) and 
stops at $u=0$ at $R=\infty$.

First, the behavior of $f=e^{u/2}$ near the origin is 
\be\label{r0}
f\approx f_{0}r^{n}\Bigl(
1-\frac{Fe^{\tilde{G}}f_{0}^{2\tilde{G}}}{8(1-\tilde{G}\tilde{M})^{2}}r^{2(1
-\tilde{G}\tilde{M})}+\cdots\Bigr).
\ee
Let $R=R_0$ be an arbitrary large number. If we choose $f_0$ sufficiently
small, then the higher order term in (\ref{r0}) can be neglected for $R\le
R_0$ and the energy of the particle ${\cal E}(R_0)$ at $R=R_0$ is given by
\be
{\cal E}(R_0)=\frac12\left(\frac{du}{dR}\right)^2+V_{eff}\Bigr|_{R=R_0}
   \simeq\frac{2n^2}{[1-\tilde{G}(n+\tilde{M})]^2R_0^2}\,.
\ee
Let us choose $R_0$ (and $f_0$, correspondingly) such that ${\cal
E}(R_0)<V_{eff}(0)$. Then since the particle energy decreases as the particle
moves due to friction, ${\cal E}(R)<V_{eff}(0)$ for $R>R_0$. In other words,
for sufficiently small $f_0$, the hypothetical particle turns back at a
point and goes to negative infinity as $R\rightarrow\infty$. Next, we
choose $f_0$ arbitrarily large and $R_0$ sufficiently small so that (\ref{r0})
is good at $R=R_0$. Then it is not difficult to show that for $R>R_0$,
\be\label{urm}
u(R)>2\ln{f_0r^n}+\frac{V'_M}{4}(R_0^2-R^2)+\frac{V'_M}{2}R_0^2\ln{R_0/R}\,,
\ee 
where $\displaystyle V'_M=\max_{-\infty<u\le0}\frac{dV_{eff}}{du}$. 
The right hand side of Eq. (\ref{urm}) has the maximum value at 
$R=R_1\equiv \left[R_0^2+\frac{4n}{V'_M[1-\tilde{G}(n+\tilde{M})]}\right]^{1/2}$
and then $u(R_1)$ satisfies, for small $R_0$, 
\be
u(R_1)>\frac{n}{1-\tilde{G}(n+\tilde{M})}
  \left(\ln\frac{4n[1-\tilde{G}(n+\tilde{M})]}{V'_M}-1\right)+2\ln f_0\,.
\ee
Therefore if we choose $f_0$ sufficiently large, then $u(R_1)>0$, {\it i.e.} the
particle goes over the hilltop of the potential. From these results, 
continuity now 
guarantees the existence of the vortex solution connecting the boundary
values, $u(0)=-\infty$ and $u(\infty)=0$, for an appropriate $f_{0}$.
This completes the proof.

The geometry of $\Sigma$ for this solution can be read from the behavior 
of the metric $b(r)$: it is singular at the origin when $\tilde{M}\neq 0$ 
and the space $\Sigma$ has an apex there due to the point particle.
As $r$ goes to infinity, the radial distance $\rho$
and the circumference $l$ diverge. The asymptotic region in terms of $R$ and 
$\theta'$ $(\theta'\equiv(1-\tilde{G}(n+\tilde{M}))\theta)$ is flat since the 
solutions approach their boundary values exponentially (see Fig. 2-(a),(b))
\be
f\approx 1-f_{\infty}K_{0}((1-\tilde{G}(n+\tilde{M}))R),
\ee
where $f_{\infty}$ is a constant determined by the proper behavior of the fields near
the origin. As shown in Fig. 2-(c), the asymptotic structure of $\Sigma$
is a cone
with deficit angle $\delta=2\pi\tilde{G}(n+\tilde{M})$.
The solution for $M=0$ has been found in Ref. \cite{LCG}.

Now we consider the case $\tilde{G}(n+\tilde{M})=1$. In this case we define
$R$ as $R=\ln r$ ($-\infty<R<\infty$) which reflects the scale 
symmetry $(r\rightarrow\lambda r)$
of the Bogomol'nyi equation (\ref{beq}) at this critical value, and then it 
takes the same
form as Eq.(\ref{New1}) with no friction term. 
Therefore in this case the hypothetical particle moves under a conservative
force only and hence the Bogomol'nyi equation can be integrated to the first
order equation
\be \label{fir}
\frac{1}{2}\left(\frac{du}{dR}\right)^{2}+V_{eff}(u) ={\cal E},
\ee
where ${\cal E}$ is a constant which is interpreted as the energy of 
the hypothetical particle which is conserved in this case. 
The particle energy ${\cal E}$ is also determined by the initial behavior
(\ref{bczero}) as
\be
\left.{\cal E}
   =\frac{1}{2}\left(\frac{du}{dR}\right)^2\right|_{R=-\infty}=2n^2\,.
\ee
In terms of classical mechanics the vortex solution is described as follows: a
hypothetical particle of unit mass with energy ${\cal E}=2n^2$ starts at 
position $u=\infty$ at time $R=-\infty$,
climbs the hill of the potential $V_{eff}$, and finally
stops at the top of hill ($u=0$) at $R=\infty$. 
For such solutions we must have $V_{eff}(0)={\cal E}=2n^2$. From the 
definition of $V_{eff}$, $V_{eff}(0)=F/\tilde{G}$ and 
the constant $F$ is determined for each $n$ as $F=2\tilde{G}n^2$.
Thus we have completely determined free parameters in this case
and obtain the solution as the form 
\be
\frac{1}{2n}\int\frac{du}{\sqrt{1-e^{-\tilde{G}(e^{u}-u-1)}}}= \int dR.
\ee
which is, unfortunately, not integrable to a closed form.
It is amusing to note that this kind of analysis is not possible in flat case.
The behavior of solution near the
origin is the same as that in Eq.(\ref{r0}), so that the space $\Sigma$ also has 
a apex at the origin for $\tilde{M}\neq0$ solutions. However, since $b(r)\sim r^{-2}$ for large $r$, 
the radial distance from the origin $\rho$ diverges but the circumference 
$l$ approaches a finite value 
$\displaystyle{l=\frac{2\sqrt{2}\pi n \sqrt{\tilde{G}}}{ev}}$ though $r$ goes 
to infinity. Then the space $\Sigma$ comprises asymptotically a cylinder as
shown in Fig. 3.

Lastly, let us consider the case that $\tilde{G}(n+\tilde{M})$ is larger 
than one. If
$n=0$, $\tilde{G}\tilde{M}>1$ and the only allowed finite energy
solution is the trivial one $u=0$ as we have seen in Eq. (\ref{mcondition}).
Suppose that there exists a solution when $\tilde{G}(n+\tilde{M})>1$ with
$n\neq0$. 
According to the similar argument given before, the space $\Sigma$ also has 
an apex if $\tilde{G}\tilde{M}$ is not zero. Since the metric $b(r)$ decreases
more rapid than $1/r^2$ for large $r$, the radial distance $\rho(r=\infty)$ is
finite and the circumference $l(r=\infty)$ vanishes. 
Then $\Sigma$ is compact and, since the Euler number given in
Eq.(\ref{Eul}) must be nonnegative, two dimensional sphere is the unique candidate. 
{}From now on let us call the point which corresponds to $r=0$ 
``the south pole" on $S^{2}$ and that which
corresponds to $r=\infty$ ``the north pole" on $S^{2}$. 
Then, at the north pole, $\phi$ does not vanish and is not well-defined;
$\phi=\phi(r=\infty)e^{in\theta}$. Therefore there is no regular solution in
this case.

\subsection*{\normalsize\bf (b) $e^{u(r=\infty)/2}=0$}
\indent\indent Now let us consider the case that the scalar field $e^{u/2}$ vanishes at
$r=\infty$ and suppose that it behaves like
$e^{u/2}\approx r^{-\varepsilon}$ for large $r$. Then the finite energy
condition from Eq. (\ref{Den}) forces $\varepsilon$ to satisfy
$\tilde{G}(n+\varepsilon+\tilde{M})>1$. Examining the asymptotic behavior of
the metric $b(r)$, we find that the radial distance  $\rho(\infty)$ and the
area $\cal A$ of the manifold are finite, and the circumference $l$ vanishes at
$r=\infty$. Therefore the space $\Sigma$ should form a two dimensional sphere $S^2$ and
it is described by our coordinate $(r,\theta)$ except the north pole where a
point particle may sit; let the mass be $M_n$. The Euler invariant given in
Eq. (\ref{eule}) then should be equal to that of $S^2$,
\be \label{seuler}
\tilde{G}(n+\varepsilon+\tilde{M}_{s}+\tilde{M}_{n})=2\,,
\ee
where $\tilde{M}_s=\tilde{M}$ and the subscript $s$ is attached as an
indication that it represents the mass of particles at the south pole. Now we
show that $\varepsilon$ can not be arbitrary by imposing the regularity
condition. From the behavior of the metric $b(r)$ at large $r$, the radial
distance $\rho(r)$ behaves near the north pole as 
$\rho-\rho(\infty)\sim r^{-\tilde{G}(n+\varepsilon+\tilde{M}_{s})+1}$. On the
other hand, if we choose a coordinate around the north pole rather than around
the south pole, we can do all the analysis we have done by replacing $M_s$ by
$M_n$. For example, regularity requires that the scalar field behaves as 
$f\sim s^n$ for $s\sim 0$, where $s$ is the radial coordinate in the new
coordinate whose origin is at the north pole; it means that
$r^{-\varepsilon}\sim s^{n}$. Also, the radial distance near the
north pole will behave as $\rho-\rho(\infty)=s^{-\tilde{G}\tilde{M}_n+1}$.
Comparing these with Eq. (\ref{seuler}), we get the consistency condition 
$r\propto s^{-1}$ and thereby $\varepsilon=n$. 

Now we discuss the existence of solutions. At first if $n=0$, there is no
nontrivial solution because $\tilde{G}\tilde{M}>1$. Next let us consider the
case $\tilde{G}(n+\tilde{M}_s)\neq1$, {\it i.e.} $\tilde{M}_s\neq\tilde{M}_n$.
Without loss of generality we may assume that $\tilde{G}(n+\tilde{M}_s)<1$.
With $\tau\equiv\ln R$ ($-\infty\le\tau\le\infty$), the Bogomol'nyi 
equation (\ref{New1}) is rewritten as 
\be
\frac{d^{2}u}{d\tau^{2}}=-e^{2\tau}\frac{dV_{eff}}{du}.
\ee
Integrating over $\tau$ from $-T$ to $T$,
\be \label{hyp}
\frac{1}{2}\left.\left(\frac{du}{d\tau}\right)^{2}\right|^{\tau=T}_{\tau=-T}
=-e^{2\tau}\left. V_{eff}\right|^{\tau=T}_{\tau=-T}+2\int^{T}_{-T}d\tau\,
e^{2\tau}V_{eff}.
\ee
{}From the behavior near the poles, $f\sim r^n$ ($r\approx 0$) and $f\sim
r^{-\varepsilon}$ ($r\rightarrow\infty$), it is easy to check that the left
hand side becomes 
\be
\frac{1}{2}\left.\left(\frac{du}{d\tau}\right)^{2}\right|^{\tau=T}_{\tau=-T}
=\frac{2}{[1-\tilde{G}(n+\tilde{M}_{s})]^{2}}\Bigl(\varepsilon^{2}
-n^{2}+O(e^{-T})\Bigr),\quad T\rightarrow\infty.
\ee
On the other hand, the first term of the right hand side in Eq. (\ref{hyp}) is
$O(e^{-T})$ while the second term goes to a positive definite and finite limit
as $T\rightarrow\infty$. Therefore, in this case, $\varepsilon^2-n^2>0$, which
is contradictory to the  aforementioned condition $\varepsilon=n$, {\it i.e.} 
there is no regular solution if $\tilde{M}_s\neq \tilde{M}_n$.

The only remaining case is that with $\tilde{M}_s=\tilde{M}_n(=\tilde{M})$, or
$\tilde{G}(n+\tilde{M})=1$. But in this case the second-order Bogomol'nyi
equation reduces to the first-order equation (\ref{fir}) and we can simply
extend the discussion below Eq. (\ref{fir}). For sphere solutions to exist the
maximum of the potential $V_{eff}(0)$ have only to be larger than the energy
of the hypothetical particle. [The condition $\varepsilon=n$ is automatically
satisfied.] That is, if $F>2n^2\tilde{G}$ for a given $n$, there always exists
a unique solution which supports two sphere. 
If we look at 
the shape of scalar field in $r$-coordinate, it resembles $n\neq0$ 
nontopological vortex solution: scalar field vanishes both at $r=0$ and at
$r=\infty$. However, since this solution comprises two 
sphere, it can be interpreted as a configuration that two vortices with
vorticity $n$ lie both at the south pole and at
the north pole, and two particles make an apex 
at each pole if $M\neq0$ (see Fig. 4). Similar to the cylinder case where
$F=2n^2\tilde{G}$, the maximum 
circumference of two sphere along the tropical line is $\displaystyle{
\frac{2\sqrt{2}n\sqrt{\tilde{G}}}{ev}}$ which is independent of $F$. 
If we regard the point particles as Planck scale strings \cite{LG} 
parallel to strings
of which the symmetry breaking scale is lower than the Planck scale, {\it e.g.}
GUT scale, it may imply a possibility of compactification of spatial 
manifold in the lower symmetry breaking scale.

\setcounter{section}{5}
\setcounter{equation}{0}
\begin{center} \section*{\large\bf V. Conclusion}
\end{center}
\indent\indent
In this paper we studied a self-dual system of (2+1) dimensional Einstein
Maxwell Higgs theory with or without external particles. 
Bogomol'nyi-type bound for the original and the dual-transformed theory
has been derived under a specific condition for the form of $\phi^{4}$
scalar potential. Then, using cylindrical symmetry ansatz, we found all 
possible soliton solutions of the equation. One type of solutions has a Higgs 
vacuum value as the boundary value and the underlying spatial manifold of 
these solutions is an asymptotic cylinder or a cone. The other, which is 
absent in the flat spacetime case, has a symmetry-restored local 
maximum value as the boundary value and the spatial
manifold constitutes two sphere with two vortices or vortex-particle 
composites. 
These solutions exist when 
the gravitational constant and the mass of external
particles satisfy the relation $\tilde{G}(n+\tilde{M})=1$. 

While the existence of cylindrically symmetric solutions of Bogomol'nyi 
equation has been rigorously demonstrated, the stability of those 
solutions, multi-soliton solutions with zero-mode counting, the existence 
of such solutions away from the Bogomol'nyi limit remain to be clarified. 
Some aspects of classical self-dual solitons were discussed using the 
dual-transformed form of the model, and
the quantum field theoretic issues such as the phase 
transition structure need further study. As the particular choice
we made for $\phi^{4}$ scalar potential was understood by a consideration of 
supersymmetric 
models in flat spacetime \cite{DeL}, a supergravity version of the model may
be interesting to investigate.

\begin{center}\section*{\large\bf Note Added}\end{center}
\ \indent
After writing this paper, we became aware of other references closely related
with the present paper. The possibility of cosmic strings of cylinder type or
2-sphere type was first proposed by Gott \cite{gott}. In Abelian Higgs model
without point particles, Linet \cite{Linet} found cylinder and sphere
configurations by solving the Bogomol'nyi equations under the choice of
parameters $n\tilde G=1$. (But he misinterpreted the sphere solution as
representing a vortex-antivortex pair; it should be interpreted as
representing a vortex-vortex pair as we have seen in Sec.\ 4.) 
Ortiz \cite{ortiz} studied the same model out of Bogomoln'yi limit 
including solutions with singularity. We thank Ortiz for bringing 
these references to our attention.

\begin{center}\section*{\large\bf Acknowledgments}\end{center}
\indent\indent
The authors would like to thank Choonkyu Lee, G. W. Gibbons, Kimyeong Lee, 
A. Polychronakos, Seyong Kim, A. Hosoya and Jooyoo Hong for helpful discussions.
This work was supported in part by the Ministry of Education, Korea and
the Korea Science and Engineering Foundation through CTP.
The work of Y. K. was supported by the Postdoctoral Fellowship of Kyung
Hee University and JSPS Fellowship under ID$\#$93033.

\def\hebibliography#1{\begin{center}\subsection*{References
}\end{center}\list
  {[\arabic{enumi}]}{\settowidth\labelwidth{[#1]}
\leftmargin\labelwidth	  \advance\leftmargin\labelsep
    \usecounter{enumi}}
    \def\newblock{\hskip .11em plus .33em minus .07em}
    \sloppy\clubpenalty4000\widowpenalty4000
    \sfcode`\.=1000\relax}

\let\endhebibliography=\endlist

\begin{hebibliography}{100}
\bibitem{Abr} A. A. Abrikosov, Zh. Eksp. Teor. Fiz. {\bf 32} (1957) 1442
$[$\spj{5} (1957) 1174$]$.
\bibitem{NO} H. B. Nielsen, and P. Olesen, \np{B61} (1973) 45.
\bibitem{Vil} A. Vilenkin, \pr{D23} (1981) 852.
\bibitem{GHV} For a review, see e.g. A. Vilenkin, \prp{121} (1985) 341;
recent developments appear in {\it The Formation and Evolution of Cosmic Strings}
edited by G. Gibbons, S. Hawking, and T. Vachaspati, (Cambridge University Press,
Cambridge, 1990).
\bibitem{DJH} A. Staruszkiewicz, Acta. Phys. Polo. {\bf 24} (1963) 735; 
S. Deser, R. Jackiw, and G. 't Hooft, \ap{152} (1984) 220; J. R. Gott III, and M.
Alpert, Gen. Rel. Grav. {\bf 16} (1984) 243; S. Giddings, J. Abbot, and
K. Kuchar, Gen. Rel. Grav. {\bf 16} (1984) 751.
\bibitem{Bog} E. Bogomol'nyi, \sjnp{24} (1976) 449.
\bibitem{HKP} J. Hong, Y. Kim, and P. Y. Pac, \prl{64} (1990) 2330; R. Jackiw,
and E. J. Weinberg, {\it ibid} {\bf 64} (1990) 2334.
\bibitem{LCG} B. Linet, \pl{A124} (1987) 240; Gen.\ Rel.\ Grav.\ {\bf20} (1988) 451;
A. Comtet, and G. W. Gibbons, \np{B299} (1988) 719.
\bibitem{Val} P. Valtancoli, \ijmp{A7} (1992) 4335; D. Cangemi, and C. Lee, \pr{D46} (1992) 4768.
\bibitem{GORS} G. W. Gibbons, M. E. Ortiz, F. Ruiz Ruiz, and T. M. Samols,
\np{B385} (1992) 127.
\bibitem{Sav} M. B. Einhorn, and R. Savit, \pr{D17} (1978) 2583; M. E.
Peskin, \ap{113} (1978) 122.
\bibitem{KL} Y. Kim, and K. Lee, Columbia University preprint CU-TP-574.
\bibitem{CCK} B. K. Chung, J.-M. Chung, and Y. Kim, preprint KHTP-93-06.
\bibitem{LG} P. Laguna, and D. Garfinkle, \pr{D15} (1989) 1011.
\bibitem{DeL} P. Di Vecchia, and S. Ferrara, \np{B130} (1977) 93.
\bibitem{gott} J. R. Gott III, Astrophys. J. {\bf 288} (1985) 422.
\bibitem{Linet} B. Linet, Class. Quantum Grav. {\bf 7} (1990) L75.
\bibitem{ortiz} M. E. Ortiz, \pr{D15} (1991) 2521.

\end{hebibliography}

\newpage
\subsection*{\vspace{-20mm}\ }
\bce\section*{\large\bf Figure Captions}
\ece
Figure 1: A typical shape of $V_{eff}(u)$. $V_{eff}(u)$ has an
asymmetric bell shape with the maximum at $u=0$.

\vspace{4mm}

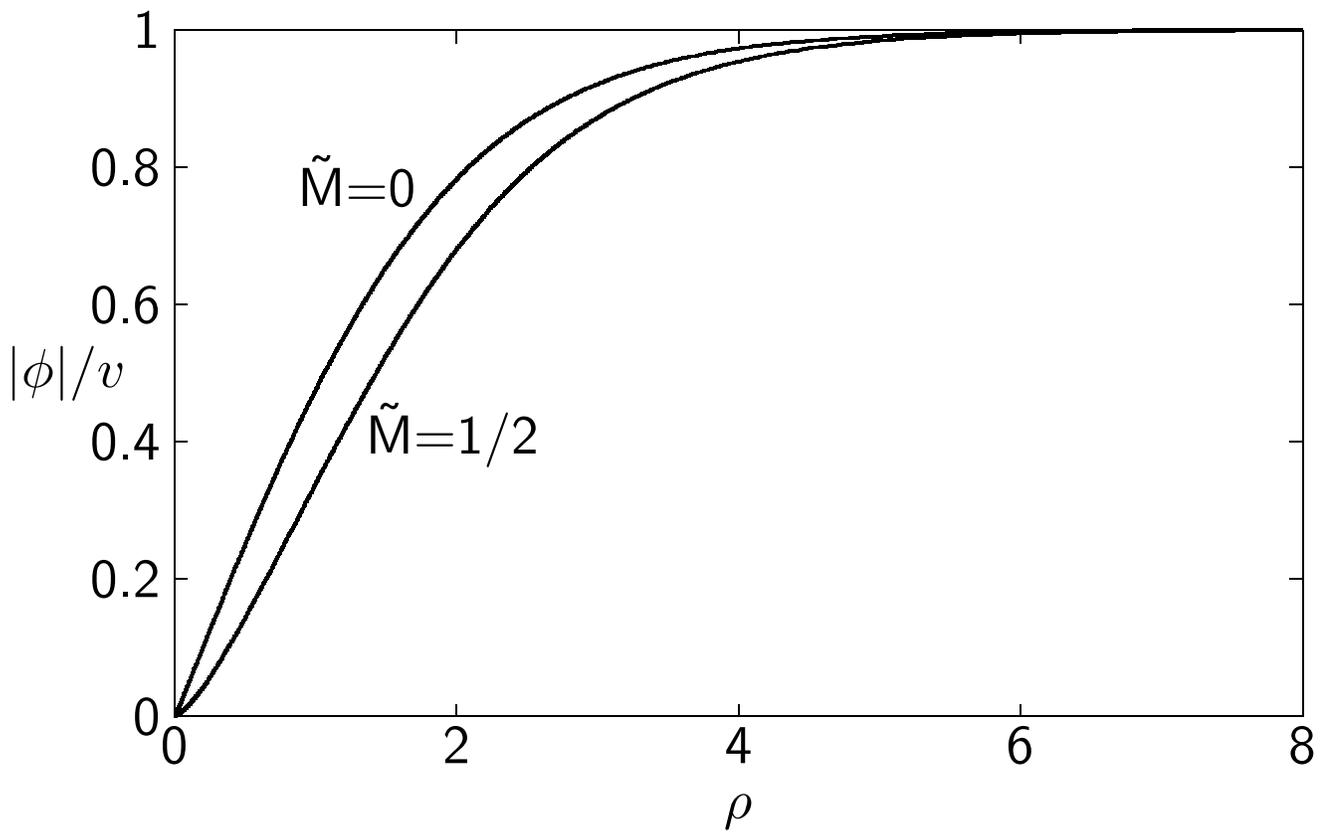
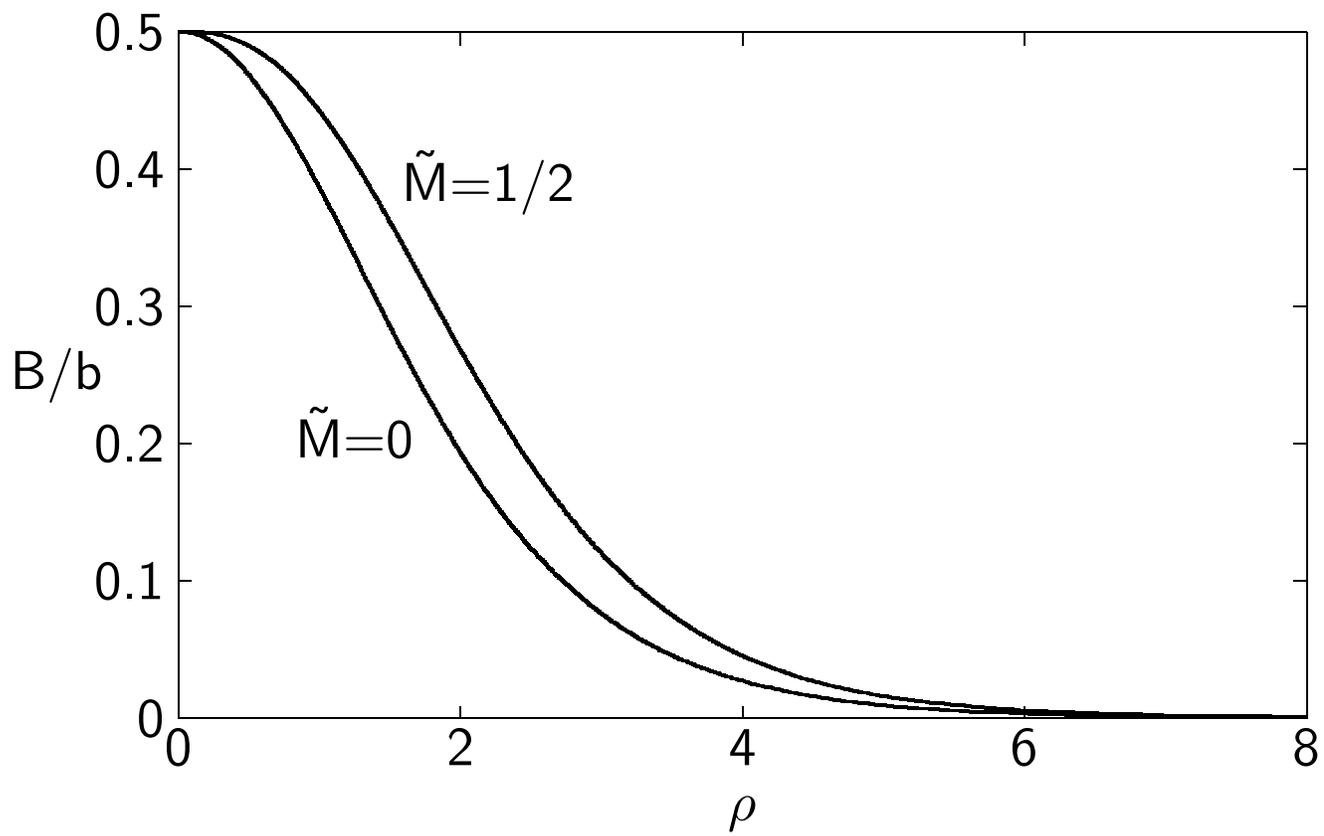
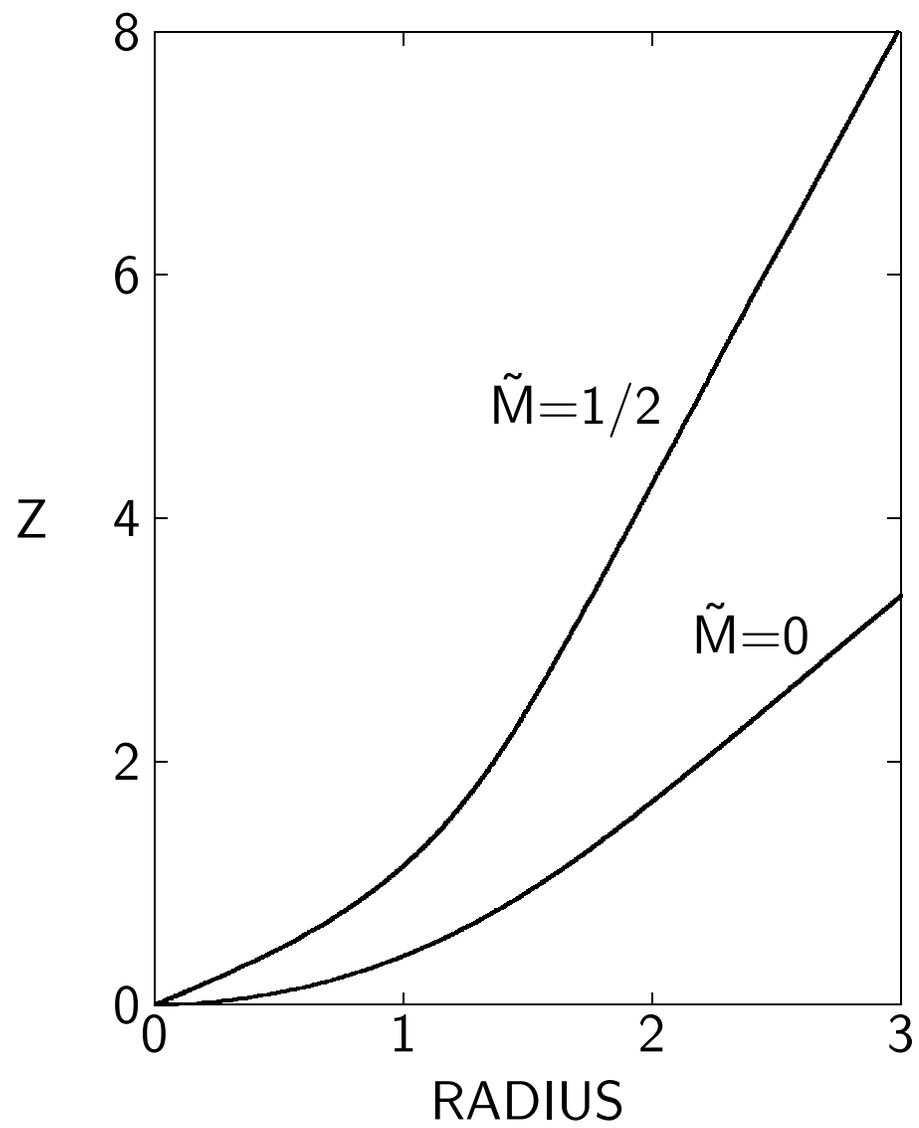
\noindent Figure 2: Plot of cylindrically symmetric solution with or without
point particles at the origin for which the underlying space form a cone at 
asymptotic region with deficit angle $\delta=2\pi\tilde G(n+\tilde M)$. 
Parameters chosen in the figures are:
$\tilde M=0$, $\tilde G=1/2$ and $F=1$ (no particle);
$\tilde M=1/2$, $\tilde G=1/2$ and $F=1$ (particles at the origin).
(a) $|\phi|/v$ vs $\rho$, (b) $B/b$ vs $\rho$ and (c) shape of the underlying
space $\Sigma$ when 
embedded in three dimensional Euclidean space with vertical coordinate 
denoted as {\sf Z}.

\vspace{4mm}

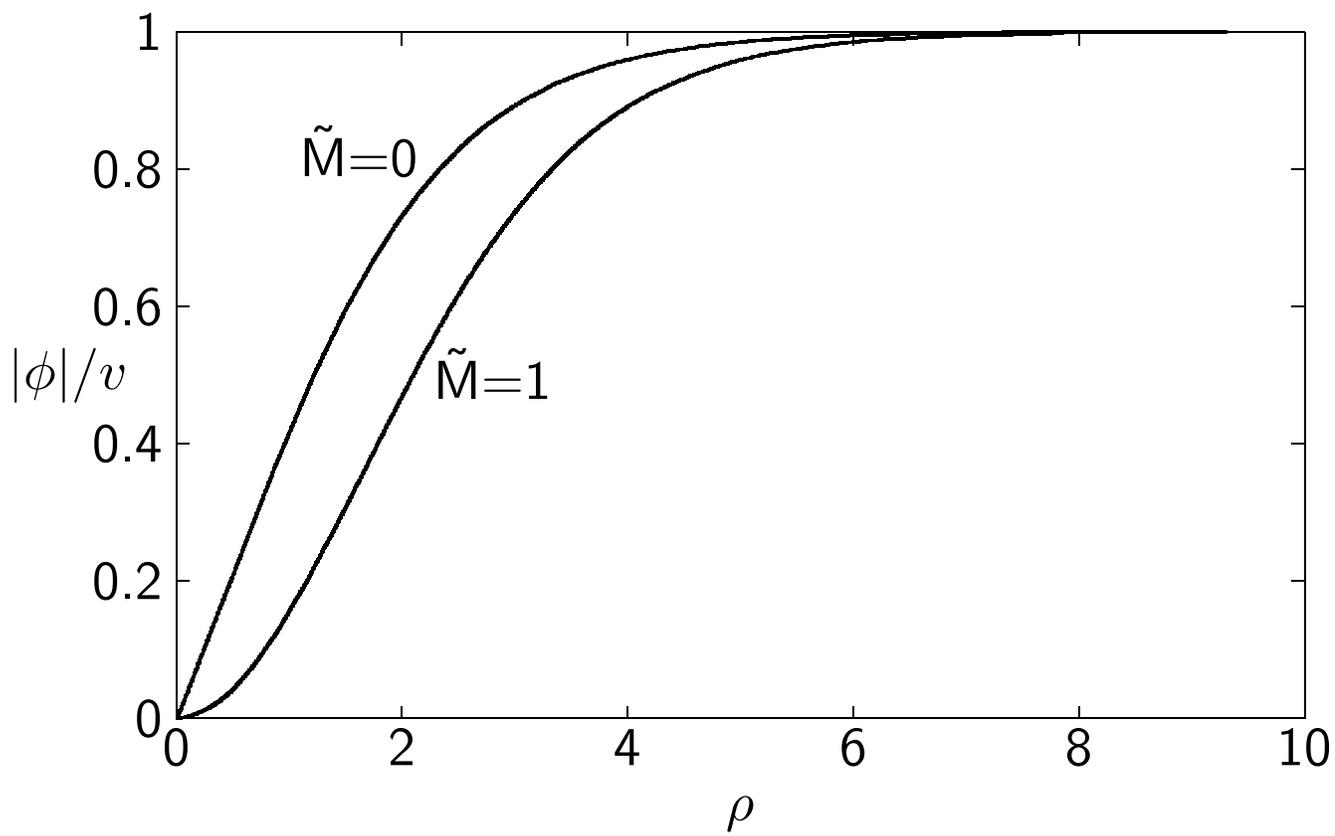
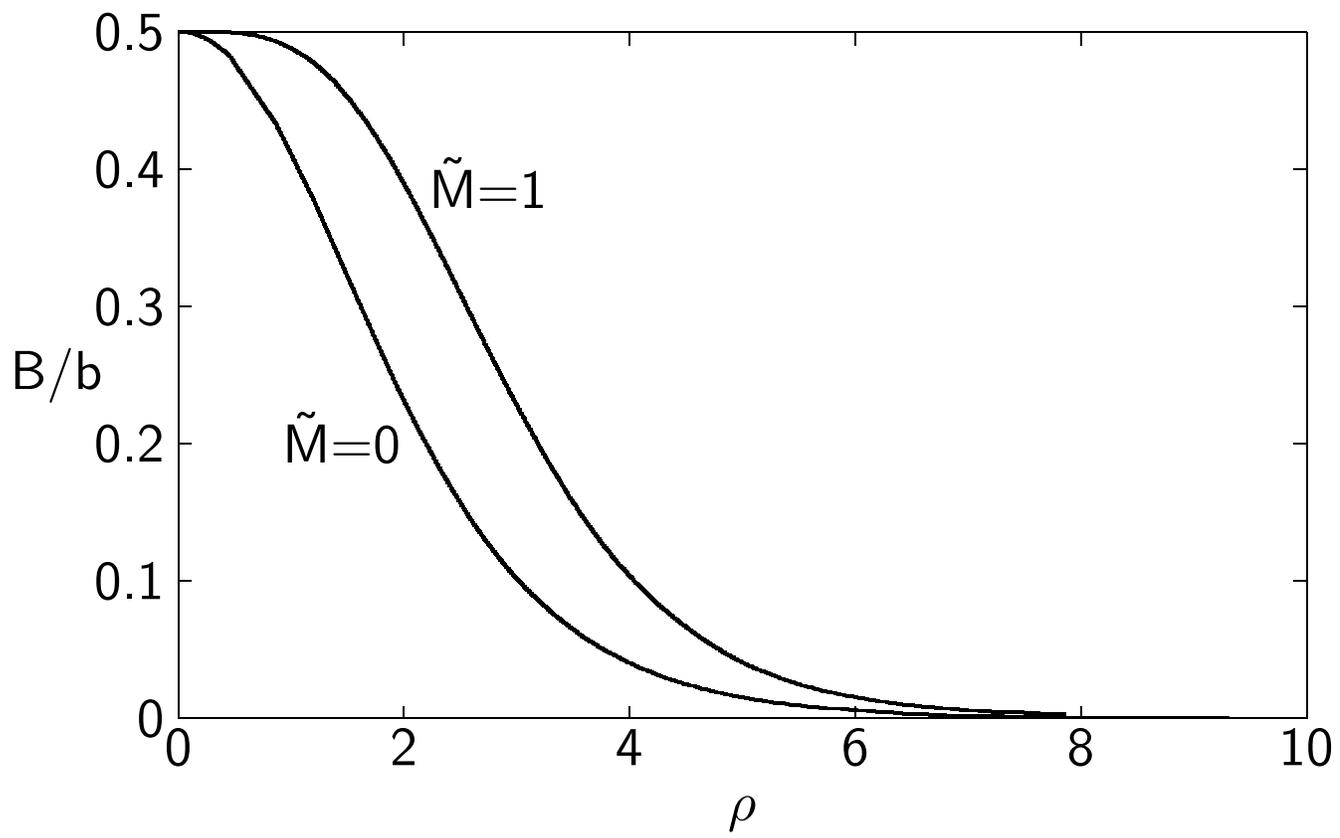
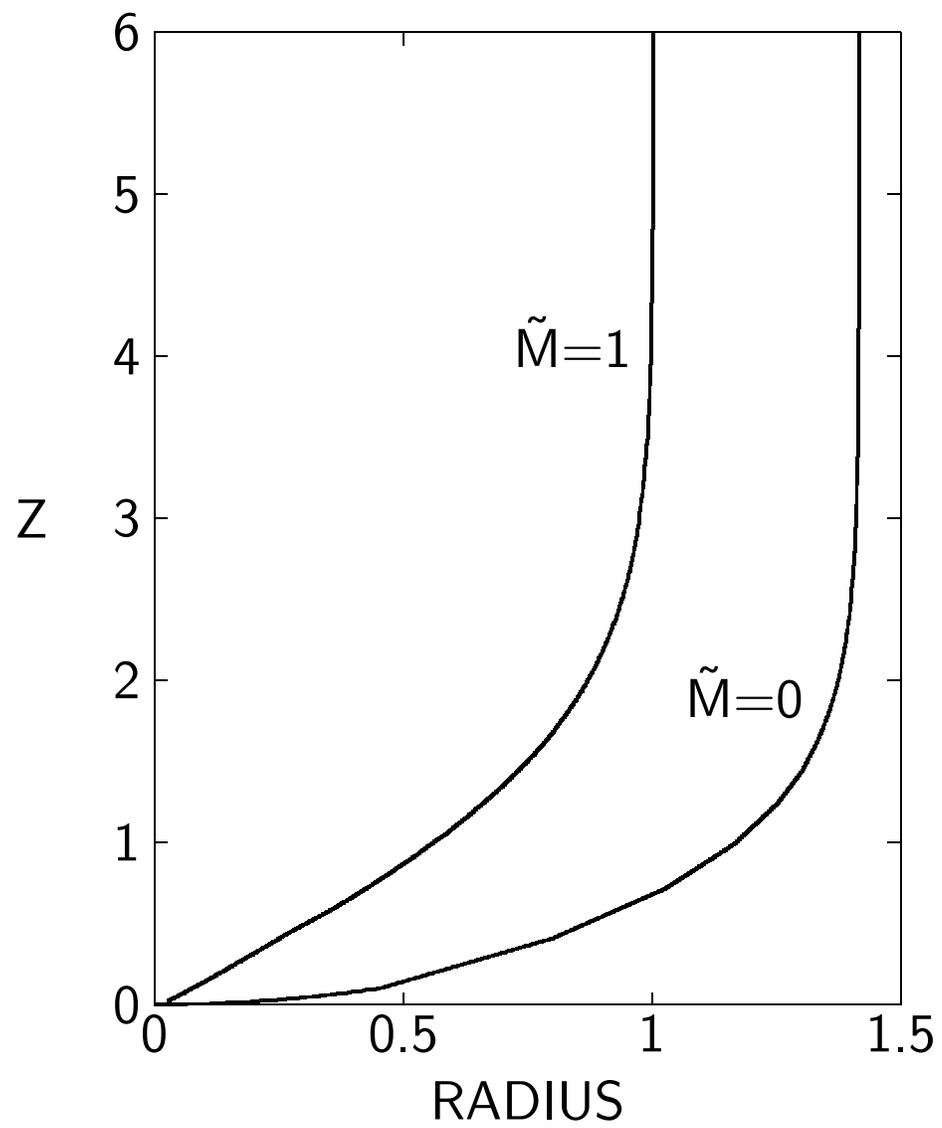
\noindent Figure 3: Plot of cylindrically symmetric solution with or without
point particles at the origin for which the underlying space form a cylinder at
asymptotic region. Parameters chosen in the figures are:
$\tilde M=0$, $\tilde G=1$ and $F=2$ (no particle);
$\tilde M=1$, $\tilde G=1/2$ and $F=1$ (particles at the origin).
(a) $|\phi|/v$ vs $\rho$, (b) $B/b$ vs $\rho$ and (c) shape of the underlying
space $\Sigma$ when
embedded in three dimensional Euclidean space with vertical coordinate
denoted as {\sf Z}.

\vspace{4mm}

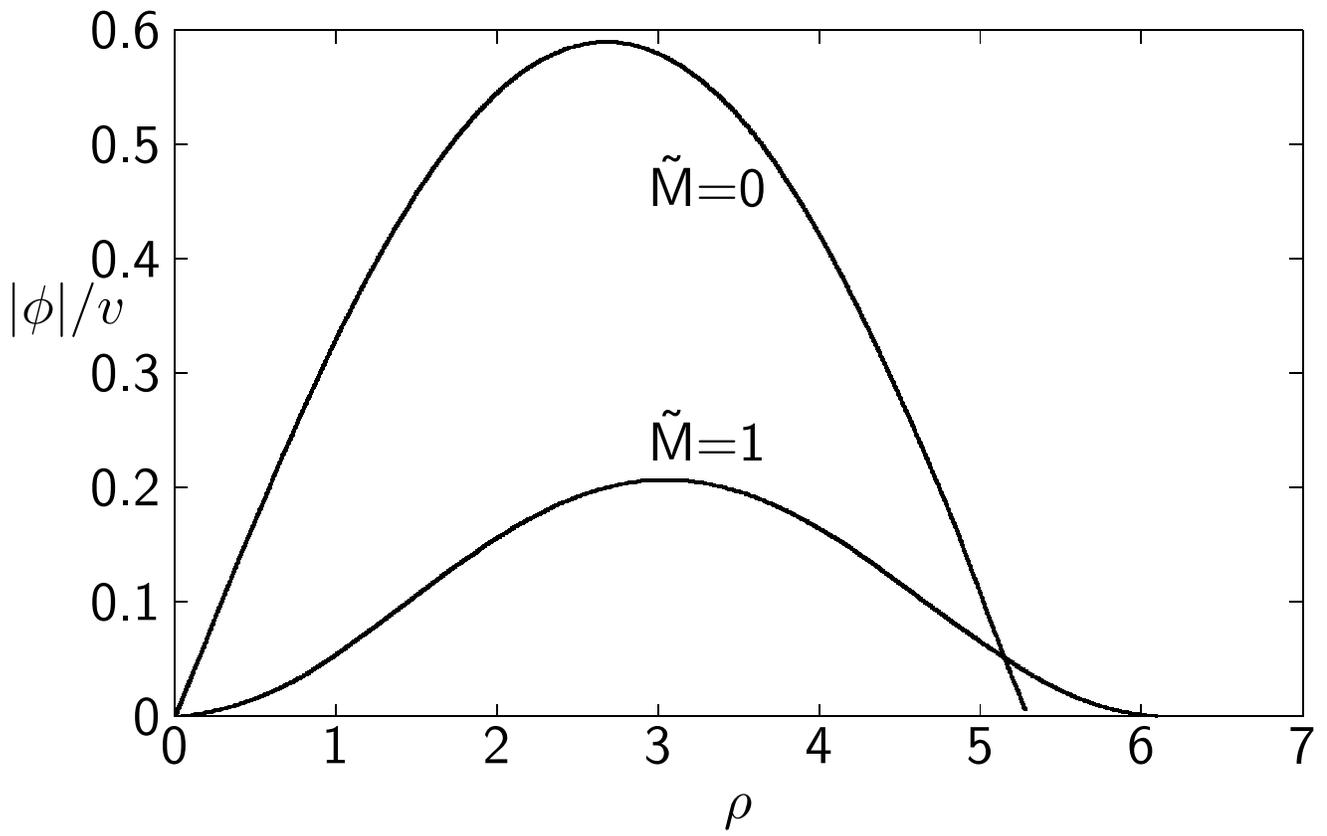
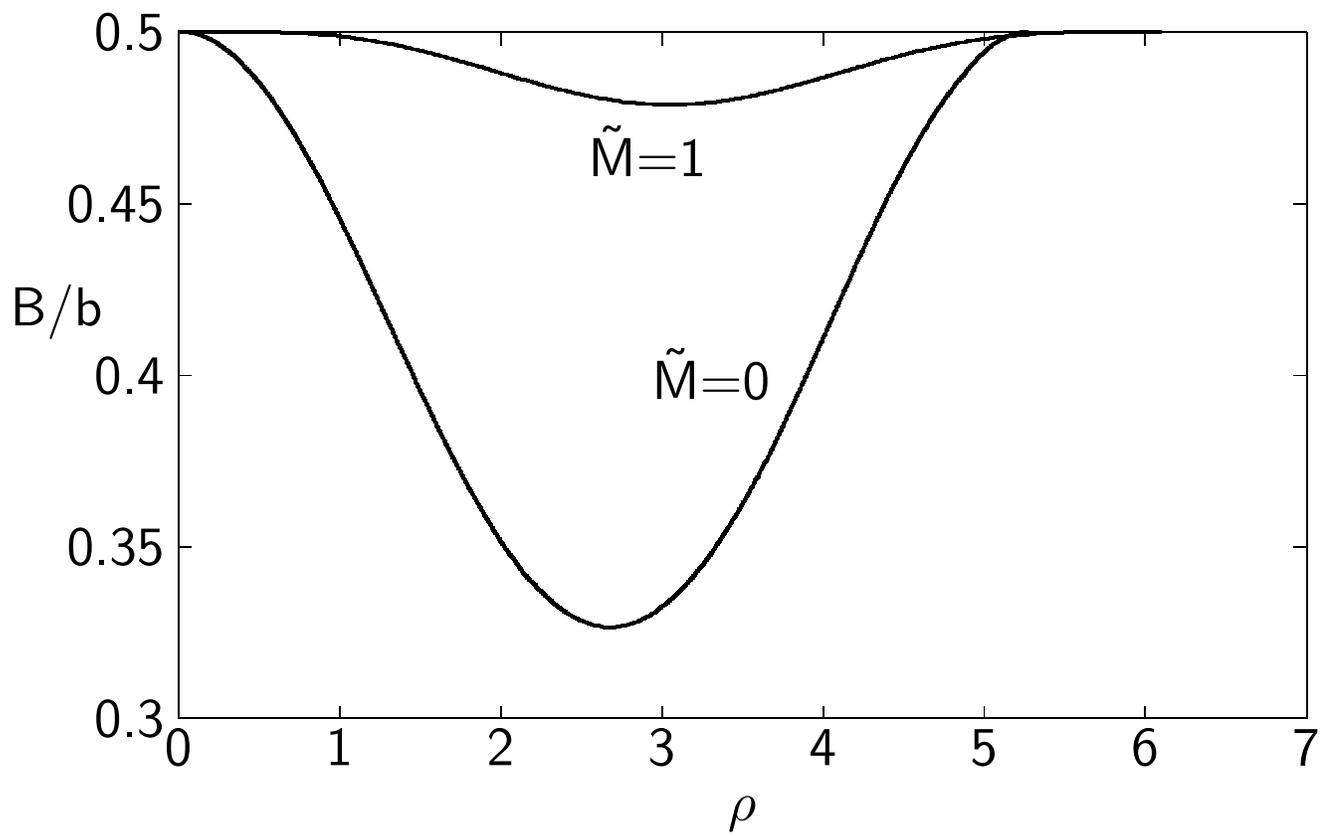
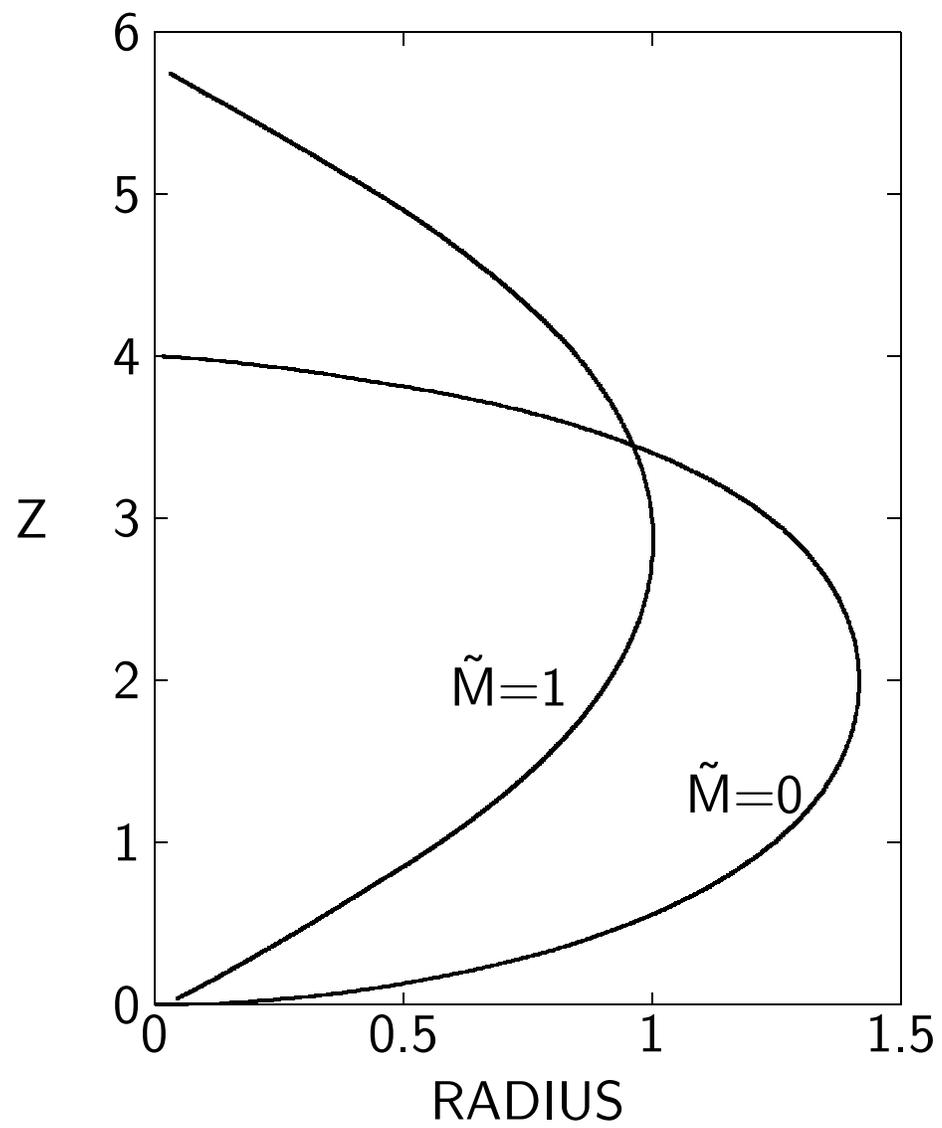
\noindent Figure 4: Plot of cylindrically symmetric solution with or without
point particles at the origin for which the underlying space form a two sphere.
Parameters chosen in figures are:
$\tilde M=0$, $\tilde G=1$ and $F=3$ (no particle);
$\tilde M=1$, $\tilde G=1/2$ and $F=3$ (particles at the origin).
(a) $|\phi|/v$ vs $\rho$, (b) $B/b$ vs $\rho$ and (c) shape of the underlying
space $\Sigma$ when
embedded in three dimensional Euclidean space with vertical coordinate
denoted as {\sf Z}.

\newpage
\input potential
\newpage
\input coned-f
\newpage
\input coned-b
\newpage
\input cone
\newpage
\input cylinderd-f
\newpage
\input cylinderd-b
\newpage
\input cylinder
\newpage
\input sphered-f
\newpage
\input sphered-b
\newpage
\input sphere

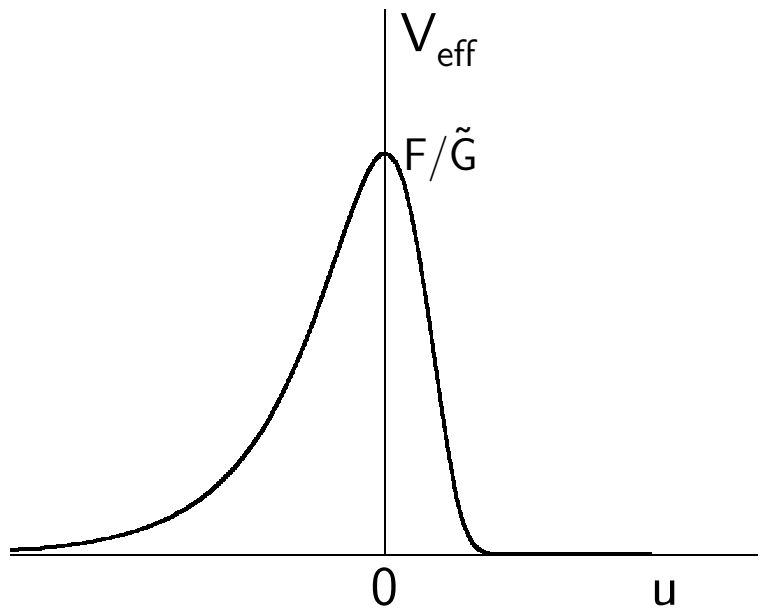
\end{document}